\documentclass[prl,twocolumn,preprintnumbers,amsfonts,amsmath,amssymb,apsa4paper]{revtex4}%
\usepackage{epsf,epsfig}
\usepackage{xcolor}
\usepackage[colorlinks=true,
            linkcolor={red!50!black},
            urlcolor={green!50!black},
            citecolor={blue!80!black}]{hyperref}
\usepackage[utf8]{inputenc}

\usepackage{booktabs}
\usepackage{graphicx}
\usepackage{dcolumn}
\usepackage{bm}
\usepackage{upgreek}
\usepackage{MnSymbol}

\def\II{{\cal I}}

\def\OO{{\cal O}}

\def\dmax{{\Delta_{\rm max}}}

\def\C{{\mathfrak C}}
\def\A{{\mathbb A}}

\newcommand{\be}{\begin{equation}}
\newcommand{\ee}{\end{equation}}
\newcommand{\beq}{\begin{equation}}
\newcommand{\eeq}{\end{equation}}
\newcommand{\ben}{\begin{displaymath}}
\newcommand{\een}{\end{displaymath}}
\newcommand{\beqa}{\begin{eqnarray}}
\newcommand{\eeqa}{\end{eqnarray}}
\newcommand{\bea}{\begin{eqnarray}}
\newcommand{\eea}{\end{eqnarray}}
\newcommand{\bean}{\begin{eqnarray*}}
\newcommand{\eean}{\end{eqnarray*}}
\newcommand{\ba}{\begin{array}}
\newcommand{\ea}{\end{array}}
\newcommand{\bi}{\begin{itemize}}
\newcommand{\ei}{\end{itemize}}

{}{}
{}{}
\def\vereq#1#2{\lower3pt\vbox{\baselineskip1.5pt \lineskip1.5pt
\ialign{$\m@th#1\hfill##\hfil$\crcr#2\crcr\sim\crcr}}}
\makeatother

\begin{document}

\title{Solving Conformal Field Theories with Artificial Intelligence}

\preprint{QMUL-PH-21-33}
\preprint{CCTP-2021-3}
\preprint{ITCP-IPP 2021/1}

\author{Gergely K\'antor}
\email{g.kantor@qmul.ac.uk}
\affiliation{Centre for Theoretical Physics, Department of Physics and Astronomy\\
Queen Mary University of London, London E1 4NS, UK}

\author{Vasilis Niarchos}
\email{niarchos@physics.uoc.gr}
\affiliation{ CCTP and ITCP, Department of Physics, University of Crete, 71303, Greece}

\author{Constantinos Papageorgakis}
\email{c.papageorgakis@qmul.ac.uk} 
\affiliation{Centre for Theoretical Physics, Department of Physics and Astronomy\\
Queen Mary University of London, London E1 4NS, UK}

\begin{abstract}
\noindent
In this paper we deploy for the first time Reinforcement-Learning algorithms in the context of the conformal-bootstrap programme to obtain numerical solutions of conformal field theories (CFTs). As an illustration, we use a soft Actor-Critic algorithm and find approximate solutions to the truncated crossing equations of two-dimensional CFTs, successfully identifying well-known theories like the 2D Ising model and the 2D CFT of a compactified scalar. Our methods can perform efficient high-dimensional searches that can be used to study arbitrary (unitary or non-unitary) CFTs in any spacetime dimension. 
\end{abstract}
\maketitle

\section{Introduction}

The generic short/large-distance behaviour of a Quantum Field Theory (QFT) is described by a Conformal Field Theory (CFT). CFTs appear in numerous physical applications, e.g.\ describing the physics of continuous phase transitions, and in many modern theoretical explorations of the non-perturbative structure of QFT. They even appear in studies of quantum gravity and black-hole physics in the context of the AdS/CFT correspondence. 

The non-perturbative solution of CFTs is a long-standing problem. One of the most promising avenues towards its resolution follows the conformal-bootstrap programme. Developed originally in the 1970s \cite{Ferrara:1973yt,Polyakov:1974gs}, the conformal bootstrap aims at the non-perturbative solution of the crossing-symmetry conditions of 4-point correlation functions. These conditions provide an infinite set of equations for an infinite set of unknowns (the CFT data), which are practically impossible, in general, to solve exactly. In modern reincarnations of the conformal bootstrap, starting with the seminal work of \cite{Rattazzi:2008pe}, one turns the problem around: Instead of searching directly for exact or approximate solutions to the crossing equations, one makes a minimal assumption about the spectrum of a unitary CFT and asks whether the crossing equations can be satisfied; if not, the assumption can be eliminated. In recent years, this approach has been implemented with great success in a variety of contexts---see e.g.~\cite{Simmons-Duffin:2016gjk,Poland:2018epd,Chester:2019wfx} for reviews---yielding allowed and disallowed regions of CFT data. On the boundary of these regions, or in `islands' of allowed regions, one can sometimes identify the data of specific CFTs. Numerical applications are facilitated by powerful linear and semi-definite programming methods, e.g.\ \cite{Simmons-Duffin:2015qma}, that have produced impressive results. A characteristic application is the 3D Ising model \cite{ElShowk:2012ht,Kos:2014bka}.

However, this implementation of the numerical conformal-bootstrap programme has limitations. Its basic modus operandi involves selecting ad hoc assumptions for a handful of free parameters, while the resulting constraints on the CFT data favour the study of special CFTs on the boundary of the allowed/disallowed regions, making it harder to study particular classes of generic theories (e.g.\ one's favourite gauge conformal field theory). Moreover, this standard approach suffers from exponential scaling on the dimensionality of the search space, which constitutes a major obstruction.

Recent literature has attacked some of these issues. Notably, \cite{Reehorst:2021ykw} introduced the `navigator-function' method: the navigator function is a globally defined, continuous, differentiable function of the CFT data, which is positive/negative in the disallowed/allowed regions. One can flow to an allowed region by minimising said function. Yet another approach to the conformal bootstrap (also applicable to non-unitary CFTs) was proposed earlier in \cite{Gliozzi:2013ysa}, where the conformal-block expansion of the crossing equations is truncated and Taylor-expanded in cross-ratio space (see also \cite{Gliozzi:2014jsa,Gliozzi:2015qsa,Gliozzi:2016cmg,Esterlis:2016psv,Hikami:2017hwv,Hikami:2017sbg,Li:2017ukc,Leclair:2018trn}). A reformulation of \cite{Gliozzi:2013ysa} as an optimisation problem was introduced in \cite{Li:2017ukc}. 
  
In this paper we introduce a novel numerical approach to the conformal bootstrap with advanced performance in high-dimensional searches that complements and/or radically improves current techniques and promises to yield previously inaccessible results. Our starting point is the algebraic system of `truncated-reduced' crossing equations of \cite{Gliozzi:2013ysa}, and their minimisation-problem reformulation in \cite{Li:2017ukc}. However, unlike \cite{Gliozzi:2013ysa,Li:2017ukc}, instead of using direct numerical methods, we take advantage of recent developments in the field of Artificial Intelligence and Machine Learning, where processes dealing with large parameter spaces have been proven to be very effective. Our preferred type of algorithms are Reinforcement-Learning (RL) algorithms \cite{sutton2018reinforcement}; RL implementations have also recently appeared in the context of String Theory \cite{Halverson:2019tkf,Ruehle:2020jrk,Harvey:2021oue,Krippendorf:2021uxu,Constantin:2021for}. By default, these do not require externally provided data for training---they learn on their own through exploration. In this work, we report very promising results based on a continuous action space RL process known as the soft Actor-Critic algorithm \cite{DBLP:journals/corr/abs-1801-01290}. 

A distinctive feature of our approach is that, instead of minimising directly the quantity of interest, we optimise a Neural Network (NN) that predicts a probability distribution. This is then sampled to yield the actual-search predictions. Unlike the direct optimisation methods that are affected by the details and complexity of the problem, the optimisation algorithms of our NNs are fixed and problem independent. In addition, our algorithms have a variable resolution of the parameter space, which explores more efficiently its global structure and is affected less by the choice of the starting point for the search. The price of our approach is that it is less `exact' and more statistical/probabilistic in nature.

We demonstrate the efficiency of the RL approach by finding approximate solutions of the truncated crossing equations that can be readily identified as CFT data for the 2D Ising model and the 2D compactified scalar. 2D CFTs were chosen as a convenient testing ground of our methods, as they include many known exactly solvable theories. We do not use, however, any of their special features (e.g.\ the structure of the Virasoro algebra) which makes our approach directly applicable to the more interesting context of higher-dimensional CFTs. In this letter, we report results of specific runs in a few special cases as a good case in point, without any systematic survey of the associated uncertainties and numerical errors. A more detailed discussion with additional explanations/examples and a preliminary discussion of the errors appears in the companion paper \cite{MLlong}.

\section{Theoretical Setup}

Unlike generic QFTs, in a CFT one can, in principle, determine all the local correlation functions of the theory through the complete knowledge of the spectrum of conformal primary operators, and their 2- and 3-point functions. The 2-point functions $\langle \OO (x_1)  \OO (x_2)\rangle$ contain information about the scaling dimensions $\Delta$ and normalisation conventions of the operators $\OO$, while the 3-point functions $\langle \OO_i(x_1) \OO_j(x_2) \OO_k(x_3)\rangle$ are fixed in terms of a single $c$-number, the 3-point function coefficient $C_{ijk}$. The latter are closely related to the coefficients $C_{ij}^k$ appearing in the Operator Product Expansion (OPE):
\beq
\label{crossaa}
\OO_i(x_1) \OO_2(x_2) = \sum_k C_{ij}^k \hat f_{ij}^k(x_1,x_2,\partial_{x_2} ) \OO_k(x_2)\;.
\eeq
$\hat f^k_{ij}$ is a differential operator that incorporates the contribution of all the conformal descendants in the conformal multiplet of the primary $\OO_k$, whose form is completely fixed by conformal symmetry. The OPE is convergent under suitable conditions and can be used to reduce higher-point correlation functions to lower-point ones. E.g., 4-point functions can be reduced to a series of squares of OPE coefficients, known as a conformal-block expansion.

We explicitly consider the case of 2D CFT. In higher dimensions some expressions become more complicated but a lot of the necessary ingredients have been significantly developed over the last decade, \cite{Simmons-Duffin:2016gjk,Poland:2018epd,Chester:2019wfx}, and can still be used to implement our approach. Following common 2D CFT conventions we call {\it primary} the operators that are highest weights of the Virasoro algebra and {\it quasi-primary} the operators that are highest weights of the $so(2,2)$ global conformal algebra. In this paper we will be using only the global part of the Virasoro algebra. Moreover, in 2D the scaling dimension $\Delta$ and spin $s$ of a quasi-primary operator $\OO$ can be expressed in terms of left- and right-moving conformal weights $h,\bar h$ respectively, as $\Delta=h+\bar h$ and $s=h-\bar h$. Unitarity requires $\Delta \geq s$. The correlator of four quasi-primary operators $\OO_i$ $(i=1,2,3,4)$ with conformal weights $(h_i,\bar h_i)$ inserted at distinct spacetime points $(z_i,\bar z_i)$ (expressed here in terms of complex coordinates in Euclidean signature) can be expanded as
\bea
\label{crossab}
&&\langle \OO_1 (z_1,\bar z_1) \OO_2 (z_2,\bar z_2) \OO_3 (z_3,\bar z_3) \OO_4(z_4,\bar z_4) \rangle
=  
\nonumber\\
&&= \frac{1}{z_{12}^{h_1+h_2} z_{34}^{h_3+h_4}} \frac{1}{\bar z_{12}^{\bar h_1+\bar h_2} \bar z_{34}^{\bar h_3+\bar h_4}}
\left( \frac{z_{24}}{z_{14}} \right)^{h_{12}} \left( \frac{\bar z_{24}}{\bar z_{14}} \right)^{\bar h_{12}}
\\
&& \times \left( \frac{z_{14}}{z_{13}} \right)^{h_{34}} \left( \frac{\bar z_{14}}{\bar z_{13}} \right)^{\bar h_{34}}
\sum_{\OO,\OO'} C_{12}^\OO g_{\OO \OO'} C_{34}^{\OO'}\, g^{1234}_{h,\bar h}(z,\bar z)
~.\nonumber
\eea
In this expression $g_{\OO \OO'}$ are 2-point function coefficients, $z_{ij}\equiv z_i-z_j$,
\beq
\label{crossac}
z = \frac{z_{12}z_{34}}{z_{13}z_{24}} ~, ~~
\bar z = \frac{\bar z_{12}\bar z_{34}}{\bar z_{13}\bar z_{24}}
\eeq
are conformally invariant combinations of the coordinates $(z_i,\bar z_i)$, known as cross-ratios, and 
\bea
\label{crossad}
g^{1234}_{h,\bar h}(z,\bar z) &=& z^h \bar z^{\bar h}  \, _2 F_1(h-h_{12},h+h_{34};2h;z)
\nonumber\\ 
&&\times\,_2 F_1(\bar h-\bar h_{12},\bar h+\bar h_{34};2 \bar h;\bar z)
\eea
are the 2D conformal blocks for an intermediate operator with conformal weights $(h,\bar h)$ \cite{Osborn:2012vt}. We are using the notation $h_{ij}=h_i - h_j$, and $_2 F_1(a,b;c;z)$ denotes the ordinary hypergeometric function. The expansion \eqref{crossab} follows from a direct application of the OPE \eqref{crossaa} simultaneously at the products $\OO_1(z_1,\bar z_1) \OO_2(z_2,\bar z_2)$ and $\OO_3(z_3,\bar z_3) \OO_4(z_4,\bar z_4)$. The intermediate sum runs over operators $\OO$, $\OO'$ with the same scaling dimension. It is convenient to define
\beq
\label{crossae}
\sum_{\OO, \OO'\, | \, \Delta_\OO = \Delta_{\OO'} = h+\bar h} C_{12}^\OO g_{\OO \OO'} C_{34}^{\OO'} \equiv \, _s \C_{h,\bar h}\;.
\eeq

Eq.\ \eqref{crossab} is known as the $s$-channel expansion. An alternative expansion arises from the $(32)-(14)$ OPEs ---the $t$-channel expansion. The equivalence of the $s$- and $t$-channel expansions yields the crossing equations
\begin{align}
\label{crossaf}
&  \sum_{h,\bar h} {_s} \C_{h,\bar h}\, g^{(1234)}_{h,\bar h}(z,\bar z) =(-1)^{(h_{41}+\bar h_{41})} \times \\
                                                                         &\frac{z^{h_1+h_2}}{(z-1)^{h_2+h_3}} \frac{\bar z^{\bar h_1+\bar h_2}}{(\bar z-1)^{\bar h_2+\bar h_3} }
\sum_{h',\bar h'} {_t} \C_{h',\bar h'}\, g^{(3214)}_{h',\bar h'}(1-z,1-\bar z)\;, \nonumber
\end{align}
which are of main interest in this paper. We note that these conformal-block sums are taken over both $(h,\bar h)$ and $(\bar h, h)$ contributions. The corresponding OPE-squared coefficients, $\C_{h,\bar h}$ and $\C_{\bar h, h}$, are generically different, but equal when the external operators are spinless. In that case, it is convenient to collect  the $(h,\bar h)$ and $(\bar h, h)$ contributions and define a single conformal block 
\bea
\label{crossafa}
&\tilde g^{(1234)}_{h,\bar h}(z,\bar z) = \frac{1}{1+\delta_{h,\bar h}} \bigg[ z^h \bar z^{\bar h} \, _2 F_1(h-h_{12},h+h_{34};2h;z)
\cr
&\, _2 F_1(\bar h-\bar h_{12},\bar h+\bar h_{34};2 \bar h;\bar z) +(z \leftrightarrow \bar z) \bigg]
~. 
\eea

To obtain a numerically tractable set of equations we {\it truncate}, assume a {\it spin-partition} and {\it reduce}. 

The truncation restricts the sums in \eqref{crossaf} to a finite number of terms, which are assumed to involve operators with scaling dimensions below an upper cutoff $\Delta_{\rm max}$. The cutoff dictates our algorithm's search window, and its presence constitutes one of the differences between our work and the truncation method of \cite{Gliozzi:2013ysa,Li:2017ukc}. The convergence properties of the conformal-block expansion \cite{Pappadopulo:2012jk} suggest that there is a sufficiently high $\Delta_{\rm max}$ above which the contribution of the truncated terms is numerically negligible. It is not a priori easy to determine this value. Moreover, if there are contributions from degenerate operators in \eqref{crossaf} only their sum counts. 

We further assume a specific assignment of spin for each contributing operator. We call this assignment the {\it spin}-{\it partition} of our setup. The total finite number of unknown CFT data is denoted $N_{\rm unknown}$. 

Finally, we need to implement a reduction from a single functional equation that depends on the continuous cross-ratio parameters $(z,\bar z)$ to a discrete set of algebraic equations. Similar reductions are performed in standard approaches of the numerical conformal bootstrap, usually by Taylor-expanding the crossing equations around the point $(z,\bar z)=(\frac{1}{2},\frac{1}{2})$. In this paper, we choose to evaluate the crossing equations \eqref{crossaf} at $N_z>N_{\rm unknown}$ different points on the $z$-plane. The sampling of these points can affect the numerics. We have observed that the sampling suggested in Ref.\ \cite{CastedoEcheverri:2016fxt} works well in our computations.

In this fashion, we obtain $N_z$ algebraic equations for $N_{\rm unknown}$ unknowns (the scaling dimensions $\Delta=h+\bar h$ of the intermediate operators in \eqref{crossaf} and their corresponding OPE-squared coefficients $\C_{h,\bar h}$), which we write as an $N_z$-dimensional vector of equations $\vec E (\vec \Delta, \vec \C) = 0$. Since we have truncated the exact crossing equations, the set $\vec E=0$ is not expected to have an exact solution. Hence, our main task is to perform an $N_{\rm unknown}$-dimensional search to find an approximate solution that minimises each component of the vector $\vec E$. We have chosen to perform this search by minimising the Euclidean norm $|| \vec E ||$ (although this choice is not unique).

When comparing different approximate solutions, we also find it convenient to define a relative measure of accuracy $\A$ that quantifies a \% violation of the truncated crossing equations. We define $\A := \frac{|| \vec E ||}{E_{\rm abs}}$, where 
\begin{align}
  \label{crossag}
  &E_{\rm abs} =\sum_{i=1}^{N_z} \Bigg[ \sum_{h,\bar h}^{\rm trunc} \bigg | {_s} \C_{h,\bar h}\, g^{(1234)}_{h,\bar h}(z_i,\bar z_i) \bigg|
\cr
&+ \bigg| z_i^{h_1+h_2} \bar z_i^{\bar h_1+\bar h_2} (z_i-1)^{-h_2-h_3} (\bar z_i-1)^{-\bar h_2-\bar h_3} \bigg|
\cr
&~~\times \sum_{h',\bar h'}^{\rm trunc} \bigg| {_t} \C_{h',\bar h'}\, g^{(3214)}_{h',\bar h'}(1-z_i,1-\bar z_i) \bigg| \Bigg]
~.
\end{align}

\begin{table}
\centering
\begin{tabular}{ | c || c | c || c | c |}
 \hline
 spin 	& analytic $\Delta$ 	& RL $\Delta$ 	& analytic $\C$  	& RL $\C$ 				\\ [0.5ex]
 \hline\hline
  0		& 4 				& 3.93	& 2.44$ \times 10^{-4}$		& 3.66$ \times 10^{-4}$			\\
  0		& 1 				& 0.99	& 0.25			& 0.25				\\
  2		& 2 				& 1.98	& 0.016		& 0.016				\\
  4		& 4 				& 3.95		& 2.2$ \times 10^{-4}$		& 2.47 $ \times 10^{-4}$			\\
  6		& 6 				& 5.97	& 1.36$ \times 10^{-5}$ 	& $0.54 \times 10^{-5}$	\\ \hline 
\end{tabular}
\caption{2D Ising CFT data for $\langle\sigma\sigma \sigma\sigma\rangle$.}
\label{table:sigma_results}
\end{table}

\section{Soft Actor-Critic algorithm}

The main components of RL algorithms are the `agent' and the `environment'. The agent explores the environment and makes decisions based on the feedback it receives. In the process, it retains its experiences and learns by optimising a NN that guides the search probabilistically. The goal of the agent is to maximise a `Reward'.

The actions of our agent are performed in a continuous space and employ a soft Actor-Critic algorithm \cite{DBLP:journals/corr/abs-1801-01290} to perform the search. The algorithm comprises an iterative process over separate steps taken by the agent. The following processes are executed in each step:
\begin{enumerate}
\item {\it Choose Action.} Each action refers to a prediction of the $N_{\rm unknown}$-dimensional vector of unknowns $(\vec \Delta, \vec \C) $.
\item {\it Implement Action in Environment.} The predicted values of the previous action are fed into the environment code.
\item {\it Observe Environment.} The reduced algebraic constraints $\vec E$ are calculated by the environment and fed back to the agent as observations.
\item {\it Obtain Reward.} The environment code generates a quantitative judgement (the `Reward') on how well the agent performed in its prediction. This reward is fed back to the agent. In our implementation the reward is $R:=-||\vec E||$. The negative sign punishes large violations of the truncated crossing equations.
\item {\it Check if Final State.} The environment checks if the agent succeeded in obtaining a better reward compared to the previous one. 
\item {\it Update Memory Buffer.} The experiences of the agent are stored in an `experience replay buffer' (containing all the information obtained by the agent in previous steps).
\item {\it Update NN.} A random sample from the memory buffer is used as training data to update the weights of the NNs. We used the ADAM optimiser in this process \cite{Kingma2015AdamAM}. In the next iteration the networks will try to predict an action with improved reward. 
\end{enumerate}

In all cases we continued running the algorithm until the relative accuracy $\A$ dropped at (or below) 0.1\% and the predictions converged to a specific profile.

\begin{table}
\centering
\begin{tabular}{ | c || c | c || c | c |}
 \hline
 spin 	& analytic $\Delta$ 	& RL $\Delta$ 	& analytic $\C$  	& RL $\C$ 			\\ [0.5ex]
 \hline\hline
  0		& 4 				& 4.07	& 1				& 1.04			\\
  2		& 2 				& 1.95	& 1				& 1.19			\\
  2		& 6 				& 5.93	& 0.1				& 0.12			\\
  4		& 4 				& 3.9	& 0.1				& 0.20			\\
  6		& 6 				& 5.93	& 0.023	 	& 0.022			\\ \hline
\end{tabular}
\caption{2D Ising CFT data for $\langle \varepsilon \varepsilon\varepsilon \varepsilon\rangle$.}
\label{table:epsilon_results}
\end{table}

\section{Results: 2D Ising model}

Our first set of results pertains to the 2D Ising model with central charge $c = \frac{1}{2}$, in which the spin primary operator $\sigma$ has conformal weights $(h,\bar h)=(\frac{1}{16},\frac{1}{16})$, and the energy-density primary operator $\varepsilon$ has $(h,\bar h) = (\frac{1}{2},\frac{1}{2})$. Their OPEs are \cite{DiFrancesco:1997nk}
\begin{align}
\label{minaa1}
&\sigma \times \sigma = [\II ] + [\varepsilon ]\;, ~
\sigma \times \varepsilon = [\sigma]\;, ~
\varepsilon \times \varepsilon = [\II ]\;,
\end{align}
where $\II$ denotes the identity operator and square brackets the family of Virasoro descendants. It is straightforward to determine the quasi-primaries in each of the conformal families appearing in \eqref{minaa1} \cite{MLlong}.

The exact crossing equation for the 4-point function $\langle \sigma(z_1,\bar z_1) \sigma(z_2,\bar z_2) \sigma(z_3,\bar z_3) \sigma(z_4,\bar z_4)\rangle $,
\begin{align}
\label{isingaa}
  {\sum_{h \geq \bar h \atop (h,\bar h)\neq (0,0)}} \hskip -.4cm\C_{h,\bar h } &\Big( |z-1|^{2\Delta_\sigma} \tilde g_{h,\bar h}^{(\sigma \sigma \sigma \sigma)}(z,\bar z) \cr
  & - |z|^{2\Delta_\sigma}  \tilde g_{h,\bar h}^{(\sigma \sigma \sigma \sigma)}(1-z,1-\bar z) \Big)\cr
& \qquad +|z-1|^{2\Delta_\sigma} - |z|^{2\Delta_\sigma}=0\;,
  \end{align}
involves intermediate operators with even spin only.
  
Setting $\Delta_\sigma = \frac{1}{8}$ for the external spin operator $\sigma$, we deployed our RL algorithm on a 10-dimensional search with $N_z = 29$ to obtain the results of Tab.~\ref{table:sigma_results}. This particular run used as input the spin-partition presented in the first column of that table with $\dmax = 6.5$. It took  12 hours on a modern laptop computer to obtain a result with relative accuracy $\A = 3.31618 \times 10^{-6}$. In Tab.~\ref{table:sigma_results} we have rounded the last digit of all the results.

Likewise, for $\langle\varepsilon(z_1,\bar z_1) \varepsilon(z_2,\bar z_2) \varepsilon(z_3,\bar z_3) \varepsilon(z_4,\bar z_4)\rangle$ the exact crossing equation is
\begin{align}
  \label{idblockaa}
  \sum_{h \geq \bar h\atop (h,\bar h)\neq (0,0)} \hskip -.4cm\C_{h,\bar h} &\Big( |z-1|^{2\Delta_\varepsilon} \tilde g_{h,\bar h}^{(\varepsilon \varepsilon \varepsilon \varepsilon)}(z,\bar z) - z^{2\Delta_\varepsilon}  \tilde g_{h,\bar h}^{(\varepsilon \varepsilon \varepsilon \varepsilon)}(1-z,1-\bar z) \Big)\cr
&\qquad+|z-1|^{2\Delta_{\varepsilon}} - |z|^{2\Delta_{\varepsilon}} =0
~.
\end{align}
Setting $\Delta_\varepsilon = 1$ for the external energy-density operator $\varepsilon$, we used our RL algorithm to obtain the results of Tab.~\ref{table:epsilon_results}. This particular run was based on the spin-partition presented in the first column of that table with $\dmax = 6.5$. We used $N_z = 29$ and it took 2 hours to reach the relative accuracy $\A = 8.62723 \times 10^{-4}$.

\begin{table}
\centering
\begin{tabular}{| c | c || c | c || c | c |}
\hline
Channel & spin 	& analytic $\Delta$ 	& RL $\Delta$ 	& analytic $\C$  & RL $\C$ 	\\ [0.5ex]
 \hline\hline
$s$ & $0$ 		& 0.4 			& 	0.39		& 1			&	0.99		\\ [1ex]
 \hline \hline
$t$ & $0$ 		 & 2 				& 2.11			& 0.01	& 0.01			\\
 &				 $1$		 & 1 				& 	0.95		& -0.1		& -0.1			\\
        & 				 $2$		 & 2 				& 	2.13		&0.005		&0.0048			\\ \hline
\end{tabular}

\caption{2D $S^1$ CFT data for $\langle V_p V_p \bar V_p \bar V_p \rangle$ with $\dmax= 2$.}
\label{table:Mom_results_1}
\end{table}

In both cases the relative accuracy at which we can satisfy the truncated crossing equations is impressive. When compared with analytic expectations, the numerical results for the scaling dimensions agree at the order of 1\%. For the OPE-squared coefficients, the agreement is equally impressive for the lower-lying operators but becomes worse for higher scaling-dimension operators that lie closer to $\dmax$. Further details about the specifics of the runs, along with several other cases, can be found in the companion paper \cite{MLlong}.

\begin{table}
\centering
\begin{tabular}{|c | c || c | c || c | c |}
 \hline
Channel & spin 	& analytic $\Delta$ 	& ML $\Delta$  &analytic $\C$ 	 		& ML $\C$ 	\\ [0ex]
 \hline\hline
$s$ & $0$ 		& 0.4 			& 0.40		& 1					& 1.01		\\
	&	 $0$ 		& 4.4 			& 4.34		& 1.23 $\times 10^{-5}$	& 0.43 $\times 10^{-5}$	\\
  	& $1$ 		& 5.4 			& 5.31		& 0					& -2.26 $\times 10^{-4}$	\\
 	&	 $2$ 		& 2.4 			& 2.41		& 3.57 $\times 10^{-3}$	& 5.49 $\times 10^{-3}$ 	\\
	&	$3$ 		& 3.4 			& 3.45		& 0					&-0.45 $ \times 10^{-5}$	\\
	&	 $4$ 		& 4.4 			& 4.41		& 1.96 $\times 10^{-3}$	& 0.28 $\times 10^{-3}$	\\
        &	 $5$ 		& 5.4 			& 5.34		& 0					&-9.98 $\times 10^{-5}$	\\ [0.4ex]
 \hline \hline
$t$ 	& 	$0$		 & 2 				& 2.00		& 1 $\times 10^{-2}$ 		& 0.57 $\times 10^{-2}$	\\
	&	$0$		 & 4 				& 4.02		& 2.5 $\times 10^{-5}$	& 4.88  $\times 10^{-4}$	\\
	&	 $1$		 & 1 				& 1.04		& -0.1				&-0.09	\\
	&	 $1$		 & 3 				& 3.05		& -5 $\times 10^{-4}$		&-2.27 $\times 10^{-2}$	\\
	&	 $1$		 & 5 				& 4.98		& -0.83 $\times 10^{-6}$   &-9.27 $\times 10^{-4}$	\\
 	&	 $2$		 & 2 				& 2.01		& 5	$\times 10^{-3}$	&1.81 $\times 10^{-3}$	\\
	&	 $2$		 & 4 				& 4.05		& 1.67 $\times 10^{-5}$	& 7.28 $\times 10^{-4}$	\\
	&	 $3$		 & 3 				& 3.03		& -1.67 $\times10^{-4}$	&-2.89 $\times10^{-4}$	\\
	&	 $3$		 & 5 				& 4.95		& -0.42 $\times 10^{-6}$ 	&-3.30 $\times 10^{-3}$	\\
	&	 $4$		 & 4 				& 3.93		&  0.42 $\times 10^{-5}$	&6.67 $\times 10^{-4}$	\\
	&	 $5$		 & 5 				& 5.04		& -8.33 $\times 10^{-8}$	&-4.36 $\times 10^{-4}$	\\ \hline
\end{tabular}

\caption{2D $S^1$ CFT data for $\langle V_p V_p \bar V_p \bar V_p \rangle$ with $\dmax= 5.5$.}
\label{table:Mom_results_3}
\end{table}

\section{Results: 2D $S^1$ CFT}

Our next set of results concern the $c=1$ 2D CFT of a compact boson $X$ with radius $R$,
 defined by the action:
\begin{align}
\label{s1analaa}
S = \frac{1}{4\pi} \int d^2 z \partial X \bar \partial X~, ~~ X \simeq X + 2\pi R
~.
\end{align}
The basic  primaries of the theory are the $U(1)$ currents
\beq
\label{s1analac}
j(z) = \frac{i}{2} \partial X(z)~, ~~ \bar j(\bar z) = \frac{i}{2} \partial X(\bar z)
\eeq
and the vertex operators
\beq
\label{s1analad}
V_{p,\bar p} (z,\bar z) = ~:e^{ipX(z) + i \bar p \bar X(\bar z)}:~, 
\eeq
where $p=\frac{n}{R}+\frac{wR}{2}$, $\bar p = \frac{n}{R}-\frac{wR}{2}$ with $n$ and $w$ the integer momentum and winding quantum numbers of the corresponding states. $j$, $\bar j$ have respectively conformal weights $(h,\bar h) = (1,0), (0,1)$ and $V_{p,\bar p}$ has $(h,\bar h) = (\frac{p^2}{2},\frac{\bar p^2}{2})$. It is again straightforward to determine the full spectrum of quasi-primary operators \cite{MLlong}.

For our first example we study 4-point functions of spinless unit-normalised vertex operators $V_p$, which are charged under the $U(1)$ symmetry of the theory: $\langle V_p(z_1,\bar z_1) V_p(z_2,\bar z_2) \bar V_p(z_3,\bar z_3) \bar V_p(z_4,\bar z_4) \rangle$. We label their scaling dimension as $\Delta_p \equiv p^2$, $p\in \mathbb R$. $\bar V_p$ denotes the complex-conjugate operator with opposite $U(1)$ charge.

The relevant crossing equation is
\begin{align}
\label{s1bb}
&\sum_{h\geq \bar h} {_s} \C_{h,\bar h} |z-1|^{2\Delta_p} \tilde g_{h,\bar h}^{(VV\bar V \bar V)}(z,\bar z) - |z|^{2\Delta_p}\cr
  &- {\hskip -0.4cm} \sum_{h'\geq \bar h' \atop (h',\bar h')\neq (0,0)} {_t} \C_{h',\bar h'} |z|^{2\Delta_p} \tilde g_{h,\bar h}^{(\bar V V V \bar V)}(1-z,1-\bar z)= 0
~.
\end{align}

In Tabs~\ref{table:Mom_results_1}, \ref{table:Mom_results_3} we have collected the expected analytic results of the $S^1$ theory, assuming that $V_p$ is a spinless operator with $\Delta_p=0.1$, together with the best results obtained by the agent. The numerical values have been truncated to the first two non-zero decimals for brevity. Already in the simplest case, corresponding to the spin partition of Tab.~\ref{table:Mom_results_1} with $\Delta_{\rm max}=2$, a run of the RL algorithm with $N_z = 30$ predicts the analytically-expected CFT data to very good accuracy. This excellent agreement persists for $\Delta$ once we incorporate operators with higher conformal dimensions, although the $\mathfrak C$ results are less accurate for smaller numbers as reported in Tab.~\ref{table:Mom_results_3} with a spin partition of $\dmax = 5.5$ and $N_z = 49$. The $N_{\rm unknown} = 8$ search of Tab.~\ref{table:Mom_results_1} took  2 hours to achieve $\mathbb A= 1.97 \times 10^{-4}$, while the $N_{\rm unknown} = 36$ run of Tab.~\ref{table:Mom_results_3}  took 2 days with $\mathbb A = 3.21 \times 10^{-5}$.

Our second correlator is the 4-point function of the holomorphic, conserved, spin-1 operator $j$,
$\langle j(z_1) j(z_2) j(z_3) j(z_4) \rangle$, yielding the crossing equation
\begin{align}
\label{s1cb}
   \sum_{h \geq \bar h\atop (h,\bar h)\neq (0,0)}\hskip -.4cm \C_{h,\bar h} &\Big( (z-1)^{2\Delta_j} g_{h,\bar h}^{(jjjj)}(z,\bar z)  - z^{2\Delta_j}  g_{h,\bar h}^{(jjjj)}(1-z,1-\bar z) \Big)\cr
    &\qquad + \frac{1}{16} \Big( (z-1)^2 - z^2 \Big)  =0
~.
\end{align}
The conserved holomorphic current $j(z)$ has spin 1 and exact scaling dimension $\Delta_j=1$. In the results reported in Tab.~\ref{table:j_results}, $\Delta_j$ was kept as one of the unknowns of the search, for which the agent determined the value $\Delta_j = 0.99\pm 0.01$. In Tab.~\ref{table:j_results} we also report statistical errors obtained by performing 10 independent runs with $N_z = 16$ and the spin-partition appearing in the first column with $\dmax = 8$. Each run took approximately 2 hours. The resulting relative accuracy was $\A = (2.14 \pm 0.08)\times 10^{-4}$. Once again, the algorithm performed very well reproducing sensibly the analytically-expected, low-lying CFT data.

\begin{table}
\centering
\begin{tabular}{ | c || c | c || c | c | c |}
 \hline
 spin 	& analytic $\Delta$ 	& RL $\Delta$ 		& analytic $\C$ & RL $\C$ 	\\ [0.5ex]
 \hline\hline
  2		& 2 				& 2.01 $\pm$ 0.02	& 2			& 1.94 $\pm$ 0.04	\\
  4		& 4 				& 3.97 $\pm$ 0.04	& 1.2			& 1.16 $\pm$ 0.04	\\
  6		& 6 				& 5.95 $\pm$ 0.04	& 0.24		& 0.22 $\pm$ 0.01	\\
  8		& 8 				& 7.98 $\pm$ 0.08	& 0.033		& 0.024 $\pm$ 0.001	\\ \hline
\end{tabular}
\caption{2D $S^1$ CFT data for $\langle j j j j \rangle$.}
\label{table:j_results}
\end{table}

\section{Outlook}

We presented examples demonstrating that the enormous recent progress in Artificial Intelligence can be harnessed to determine data in generic CFTs with very good accuracy. Further examples and more detailed discussion can be found in \cite{MLlong}. Some CFT features need to guide the search, e.g.\ some prior information about the structure of the spectrum can be used to inform the spin-partition and search windows. We believe that our approach will be extremely useful in situations where a CFT can be first identified (solved) in a parametrically convenient regime. In such cases, the perturbative information can be used to set up a spin-partition that informs searches with gradually changing parameters from weak to strong coupling. We expect that in the near future the synergy of our methods with other existing analytic and numerical approaches will lead to a plethora of new non-perturbative results in CFT. Supersymmetric CFTs are an immediate target for further formal developments. 

In order to better develop our RL algorithm, it is also imperative to better understand the various sources of error: analytic errors stemming from the truncation of the conformal-block expansion, the $z$-point sampling, statistical errors, the choice of reward function, as well as the particular choice of the learning algorithm that we used.

\vspace{0.3cm}
\section{Acknowledgements}
We would like to thank P.~Agarwal, E.V.G.~Christiansen,  S.~Kousvos and B.~van Rees for useful discussions and comments. We acknowledge financial support by the Royal Society Grant No. URF\textbackslash R\textbackslash 180009 and the STFC Grant No. ST/P000754/1.



\bibliography{machine_learning}
\bibliographystyle{apsrev}
\end{document}